\documentclass[10pt,conference]{IEEEtran}
\usepackage{pstricks}
\usepackage{graphicx}
\usepackage{amsmath,amssymb}
\usepackage{color}
\input colordvi

\def\Blue#1{\Color{0.70 0 0 0.2}{#1}}
\def\Blue2#1{\Color{0.6 0.3 0.5 0.2}{#1}}

\newcommand{\defn}{\stackrel{\Delta}{=}}

\newcommand{\bu}{{\mathbf{u}}}

\newcommand{\bv}{{\mathbf{v}}}
\newcommand{\bz}{{\mathbf{z}}}
\newcommand{\bx}{{\mathbf{x}}}
\newcommand{\bs}{{\mathbf{s}}}

\newcommand{\hbu}{\hat{{\mathbf{u}}}}
\newcommand{\be}{{\mathbf{e}}}
\newcommand{\by}{{\mathbf{y}}}

\newcommand{\bY}{{\mathbf{Y}}}

\newcommand{\bU}{{\mathbf{U}}}

\newtheorem{Example}{Example}

\begin{document}

\title{Channel combining and splitting for cutoff rate improvement}
\author{\authorblockN{Erdal Ar{\i}kan}
\authorblockA{Electrical-Electronics Engineering Department\\
Bilkent University, Ankara, 06800, Turkey\\
Email: arikan@ee.bilkent.edu.tr}}
\maketitle

\begin{abstract}
The cutoff rate $R_0(W)$ of a discrete memoryless channel (DMC) $W$ is often used as a figure of merit, alongside the channel capacity $C(W)$.
Given a channel $W$ consisting of two possibly correlated subchannels $W_1$, $W_2$,
the capacity function always satisfies $C(W_1)+C(W_2) \le C(W)$, while there are examples for which
$R_0(W_1)+R_0(W_2) > R_0(W)$.
This fact that cutoff rate can be ``created'' by channel splitting was noticed by Massey in his study of an optical modulation system modeled as a $M$'ary
erasure channel.
This paper demonstrates that similar gains in cutoff rate can be achieved for general DMC's by methods of channel combining and splitting.  
Relation of the proposed method to Pinsker's early work on cutoff rate improvement and to Imai-Hirakawa multi-level coding are also discussed.
\end{abstract}

\section{Introduction}
Let $W$ be a DMC with input alphabet ${\cal X}$, output alphabet ${\cal Y}$, and transition probabilities $W(y|x)$.
Let $Q$ be a probability distribution on ${\cal X}$, and define the functions
\begin{align*}
E_0(\rho,Q,W) = -\log \sum_y \left[\sum_x Q(x) W(y|x)^\frac{1}{1+\rho}\right]^{1+\rho}
\end{align*}
where $\rho \ge 0$ (all logarithms are to the base 2 throughout), and
\begin{align*}
E_r(R,Q,W) = \max_{0\le \rho\le 1} [E_0(\rho,Q,W)-\rho R]
\end{align*}
where $R\ge 0$. 
The {\sl random-coding exponent} is given by
\begin{align*}
E_r(R,W) = \max_Q E_r(R,Q,W)
\end{align*}
Gallager \cite[Theorem 5.6.2]{Gallager} shows 
that the probability of ML (maximum-likelihood) decoding error $\overline{P}_e$ over a $(N,2^{NR},Q)$ block code ensemble is 
upperbounded by $2^{-NE_r(R,Q,W)}$.
A $(N,2^{NR},Q)$ block code ensemble is one where each letter of each codeword is chosen independently from distribution $Q$.
Gallager shows that the exponent $E_r(R,W)$ is positive for all rates $0\le R< C$, where $C$ is the channel capacity.
The channel {\sl cutoff rate\/} is defined as $R_0(W) \defn \max_{Q} E_0(1,Q,W)$ and equals the random coding exponent
at rate $R=0$, {\sl i.e.} $R_0(W)=E_r(0,W)$.

Gallager's ``parallel channels theorem'' \cite[p.~149]{Gallager} states that 
\begin{align*}
E_0(\rho,W_1\otimes W_2) = E_0(\rho,W_1) + E_0(\rho,W_2)
\end{align*}
where  $W_1:{\cal X}_1\to {\cal Y}_1$ and $W_2:{\cal X}_2\to {\cal Y}_2$ are any two DMC's,
$W_1\otimes W_2$ denotes a DMC $W:{\cal X}_1\times {\cal X}_2 \to {\cal Y}_1\times {\cal Y}_2$
with transition probabilities $W(y_1,y_2|x_1,x_2)=W_1(y_1|x_1)W_2(y_2|x_2)$ for all $(x_1,x_2)\in {\cal X}_1\times {\cal X}_2$ and 
$(y_1,y_2)\in{\cal Y}_1\times {\cal Y}_2$.
This theorem implies that $E_0(\rho,W^{\otimes n}) = n E_0(\rho,W)$ and hence 
$E_r(nR,W^{\otimes n})=nE_r(R,W)$.
This is a {\sl single-letterization} result stating that the random-coding exponent cannot be improved by considering
ensembles where codewords are made up of super-symbols chosen from an arbitrary distribution $Q_n$ on blocks of $n$ channel inputs.

\subsection{Massey's example}
The independence of channels $W_1$ and $W_2$ is crucial in the parallel channels theorem;
if they are correlated then equality may fail either way.
Massey \cite{Massey81} made use of this fact to gain a coding advantage in the context of an optical communication system.
Massey's idea is illustrated in the following example; this same example was also discussed in \cite{Gallager85}.
\begin{Example}[Massey \cite{Massey81}]\label{Ex:bec2}
Consider the quaternary erasure channel (QEC), 
$W:{\cal X}_1\times {\cal X}_2\to {\cal Y}_1\times {\cal Y}_2$
where ${\cal X}_1={\cal X}_2=\{0,1\}$, ${\cal Y}_1={\cal Y}_2=\{0,1,?\}$, and
\begin{align*}
W(y_1y_2|x_1x_2) = \left\{ \begin{array}{cc} 1-\epsilon, &\quad y_1y_2=x_1x_2 \\
\epsilon, & \quad y_1y_2=??
\end{array}
\right.
\end{align*}
where $0\le \epsilon\le 1$ is the erasure probability.
The QEC $W$ can be decomposed into two BEC's (binary erasure channels): $W_i:{\cal X}_i\to {\cal Y}_i$, $i=1,2$.
In this decomposition, a transition $(x_1,x_2)\to (y_1,y_2)$ over the QEC 
is viewed as two transitions, $x_1\to y_1$ and $x_2\to y_2$, taking place on the respective component channels, with
\begin{align*}
W_i(y_i|x_i) = \left\{ \begin{array}{cc} 1-\epsilon, & y_i=x_i \\
\epsilon, & y_i=?
\end{array}
\right.
\end{align*}
These BEC's are fully correlated in the sense that an erasure occurs either in both or in none.

Humblet \cite{Humblet} gives the random-coding exponent for the $M$'ary erasure channel (MEC) as follows.
\begin{align}
\label{MECexponent}
E_{r}(R,\mbox{MEC}) = \left\{ \begin{array}{cc} D\big(1-\frac{R}{\log M}\,||\,\epsilon\big), & R_{c} \le R \le C \\
R_0 - R, & 0\le R \le R_{c}
\end{array}
\right.
\end{align}
where $D(\delta||\epsilon) =\delta \log(\delta/\epsilon) + (1-\delta)\log\big[(1-\delta)/(1-\epsilon)\big]$,
$C = (1-\epsilon)\log M$ is the capacity, $R_{c} = C/[1+(M-1)\epsilon]$ is the {\sl critical rate,\/} and 
$R_0 = \log M - \log [1+(M-1)\epsilon]$ is the cutoff rate.
Fig.~\ref{qecbecexponents} shows the random-coding exponents for the QEC and the BEC with $\epsilon=0.25$. 
It is seen from the figure that 
\begin{align}
\label{QECimprovement}
E_r(R,W) < E_r(R/2,W_1)+ E_r(R/2,W_2)
\end{align}
In fact for rates $R>R_{c}(W)=2(1-\epsilon)/(1+3\epsilon)$, the exponent is doubled by splitting: $E_r(R/2,W_1)+E_r(R/2,W_2)=2E_r(R,W)$.
Also, $C(W)=C(W_1)+C(W_2)$, i.e., the capacity of the QEC is not degraded by splitting it into BEC's.

\begin{figure}[hbt]
\begin{center}
\resizebox{!}{!}{
\includegraphics*[width=\columnwidth]{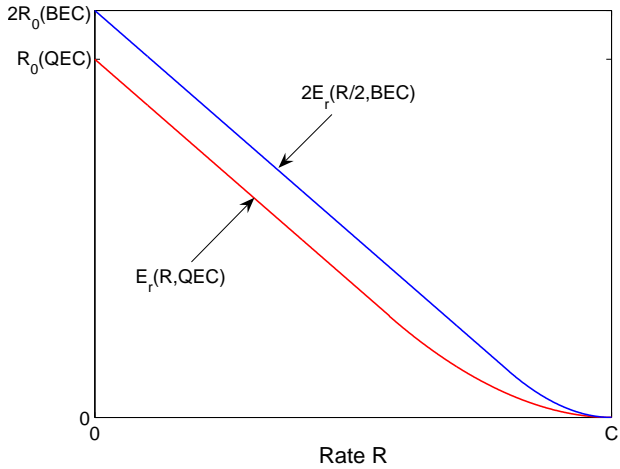}
}
\end{center}
\caption{Random-coding exponents for QEC and BEC.}
\label{qecbecexponents}
\end{figure}
Instead of direct coding of the QEC $W$, Massey suggested applying
independent encoding of the component BECs $W_1$ and $W_2$, ignoring the 
correlation between the two channels.
The second alternative presents significant advantages with respect to 
(i) reliability-complexity tradeoff in ML decoding,
and (ii) the cutoff-rate criterion.

{\sl Reliability-complexity tradeoff.}
Consider block coding on the QEC using a $(N,2^{NR},Q)$ code ensemble where $Q$
is uniform,
so that $E_r(R,W)=E_r(R,Q,W)$ for all $R$.
The ML decoding complexity $\chi$ is proportional to the number of codewords, $\chi \cong 2^{NR}$.
The reliability is given by
$\overline{P}_e \cong 2^{-NE_r(R,W)}$.

Next, consider ML decoding over the two subchannels $W_1$ and $W_2$, using
independent $(2N,2^{NR},Q')$ ensembles,
where $Q'$ is uniform.
Then, $E_r(R,\text{BEC})=E_r(R,Q',\text{BEC})$, and
the ML complexity and reliability figures are
$\chi_1 + \chi_2 \cong  2^{NR}$ and 
$\overline{P}_{e,1} +  \overline{P}_{e,2} \cong 2^{-2NE_r(R/2,\text{BEC})}$.
Thus, for the same order of complexity, the second alternative offers higher reliability due
to inequality~\eqref{QECimprovement}.

{\sl The cutoff rate criterion.}
One reason for considering the cutoff rate as a figure of merit for comparing the two coding alternatives in Massey's example
is due to its role in {\sl sequential decoding},
which is a decoding algorithm for tree codes invented by 
Wozencraft \cite{Wozencraft}.
Sequential decoding can be used to achieve arbitrarily reliable communication on any DMC $W$ 
at rates arbitrarily close to $R_0(W)$ 
while keeping the average computation per decoded digit bounded by a constant that depends on the code rate, the channel $W$, but not on the 
desired level of reliability.
Sequential decoding applied directly to the QEC can achieve 
$R_0(\text{QEC})= 2-\log(1+3\epsilon)$.
If instead, one applies independent coding and sequential decoding on the component channels,
one can achieve a sum rate of $2R_0(\text{BEC})=2[1-\log(1+\epsilon)]$, which
exceeds $R_0(\text{QEC})$ for all $0 < \epsilon < 1$, as shown 
in Fig.~\ref{CCRQEC2BEC}.
The figure shows that Massey's method bridges the gap between the cutoff rate and the capacity of the
QEC significantly.

Apart from its significance in sequential decoding, the cutoff rate serves as a one-parameter gauge of
the channel reliability exponent.
Since $R_0(W)$ is the vertical axis intercept of the $E_r(R,W)$ vs. $R$ curve, i.e., $R_0(W)=E_r(0,W)$,
an improvement in the cutoff rate is usually accompanied by an improvement in the entire random-coding exponent.
For a more detailed justification of the use of cutoff rate as a figure of merit for a communication system, we refer
to \cite{WozencraftKennedy}, \cite{Massey74}. 

\begin{figure}[hbt]
\begin{center}
\resizebox{!}{!}{
\includegraphics*[width=\columnwidth]{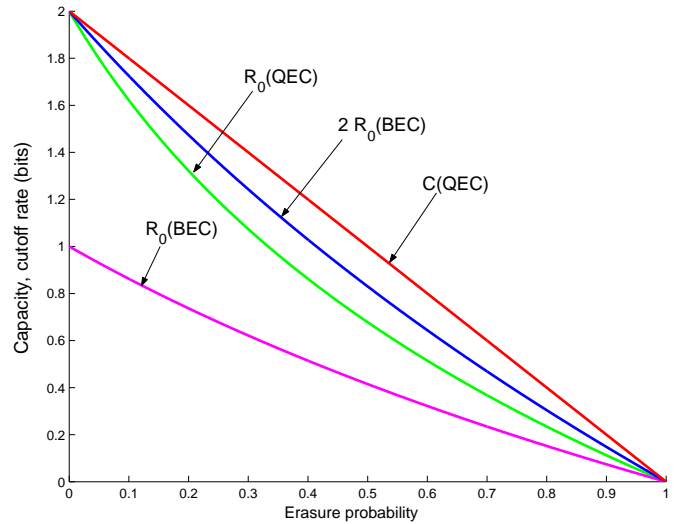}
}
\end{center}
\caption{Capacity and cutoff rate for the splitting of a QEC.}
\label{CCRQEC2BEC}
\end{figure}
\end{Example}

\subsection{Outline}
This paper addresses the following questions raised by Massey's example.
Can {\sl any} DMC be split in some way to achieve coding gains
as measured by improvements in the ML reliability-complexity tradeoff or in the cutoff rate? 
And, if so, what are the limits of such gains? 

We address these questions in the framework of coding systems that consist of three elements: (i) channel combining, (ii) input relabeling, and (iii) channel splitting.
In Massey's example there is no channel combining; a given channel is simply split into subchannels.
However, in general, it turns out that it is advantageous to combine multiple copies of a given channel prior to splitting.
Input relabeling exists in Massey's example: the inputs of the QEC which would normally be labeled as $\{0,1,2,3\}$ are instead labeled
as $\{00,01,10,11\}$.
Channel splitting is achieved in Massey's example by complete separation of both the encoding and the decoding tasks on the subchannels.
In this paper, we keep the condition that the encoders for the subchannels be independent but admit {\sl successive cancelation} or {\sl multi-level} 
type decoders where each decoder communicates its decision to the next decoder in a pre-fixed order.
In this sense, our results have connections with Imai-Hirakawa multi-level coding scheme \cite{Imai77}.

The main result of the paper is the demonstration of some very simple techniques by which significant cutoff rate improvements can be obtained
for the BEC and the BSC (binary symmetric channel).
The methods presented are readily applicable to a larger class of channels.

\section{Channel combining and splitting}\label{sec:generalframework}

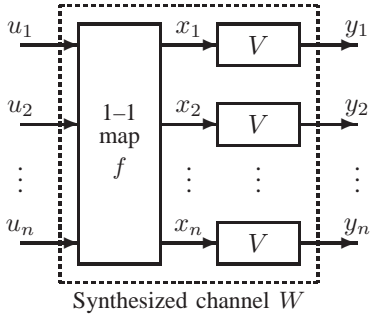
\begin{figure}[hbt]
\begin{center}
\unitlength 1.5pt
\begin{picture}(100,80)(0,0)
\thicklines
\put (35,53){\makebox(0,0)[c]{\small 1--1}}
\put (35,46){\makebox(0,0)[c]{\small map}}
\put (35,39){\makebox(0,0)[c]{\small $f$}}
\put (25,15){\framebox(20,60)[c]{}}
\put (60,65){\framebox(20,10)[c]{$V$}}
\put (60,45){\framebox(20,10)[c]{$V$}}
\put (60,15){\framebox(20,10)[c]{$V$}}
\put (10,70){\vector(1,0){15}}
\put (10,50){\vector(1,0){15}}
\put (10,20){\vector(1,0){15}}
\put (45,20){\vector(1,0){15}}
\put (45,50){\vector(1,0){15}}
\put (80,50){\vector(1,0){15}}
\put (45,70){\vector(1,0){15}}
\put (80,70){\vector(1,0){15}}
\put (80,20){\vector(1,0){15}}
\put (10,22){\makebox(0,0)[bc]{$u_n$}}
\put (10,52){\makebox(0,0)[bc]{$u_2$}}
\put (10,72){\makebox(0,0)[bc]{$u_1$}}
\put (52.5,52){\makebox(0,0)[bc]{$x_2$}}
\put (52.5,72){\makebox(0,0)[bc]{$x_1$}}
\put (52.5,22){\makebox(0,0)[bc]{$x_n$}}
\put (95,72){\makebox(0,0)[bc]{$y_1$}}
\put (95,52){\makebox(0,0)[bc]{$y_2$}}
\put (95,22){\makebox(0,0)[bc]{$y_n$}}
\put (10,38){\makebox(0,0)[c]{$\vdots$}}
\put (95,38){\makebox(0,0)[c]{$\vdots$}}
\put (70,38){\makebox(0,0)[c]{$\vdots$}}
\put (52.5,38){\makebox(0,0)[c]{$\vdots$}}
\put (20,10){\dashbox(65,70){}}
\put (52.5,5){\makebox(0,0)[c]{\small Synthesized channel $W$}}
\end{picture}
\caption{Channel combining and input relabeling.}
\label{Fig:generalmethod}
\end{center}
\end{figure}

In order to seek gains as measured by the cutoff rate, we will consider DMCs of the form
$W:{\cal X}^n \to {\cal Z}$ for some integer $n\ge 2$, obtained
by combining $n$ independent copies of a given DMC $V:{\cal X}\to {\cal Y}$, as shown
in Fig.~\ref{Fig:generalmethod}.
An essential element of the channel combining procedure is a bijective 
function $f: {\cal X}^n \to {\cal X}^n$
that relabels the inputs of $V^{\otimes n}$ (the channel that consists of $n$ independent copies of $V$).
The resulting channel is a DMC $W:{\cal X}^n \to {\cal Z}\defn {\cal Y}^n$ such that 
$W(z|u_1,\ldots,u_n)=\prod_{i=1}^n V(y_i|x_i)$
where $(x_1,\ldots,x_n)=f(u_1,\ldots,u_n)$, $(u_1,\ldots,u_n)\in{\cal X}^n$, $z=(y_1,\ldots,y_n)\in {\cal Z}$.

We will regard $W$ as an $n$-input multi-access channel where each 
input is encoded independently by a distinct user.
The decoder in the system is a successive-cancelation type decoder
where each decoder feeds its decision to the next decoder; and, there is only one pass
in the algorithm.
We will refer to such a coding system a {\sl multi-level} coding system using the terminology of \cite{Imai77}.
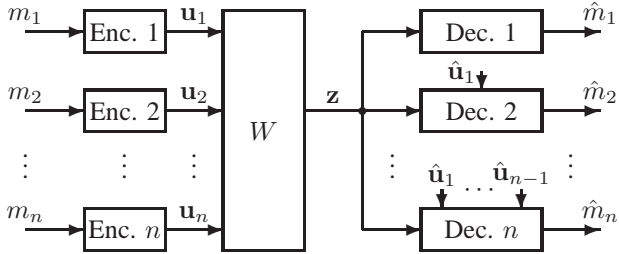
\begin{figure}[htb]
\begin{center}
\unitlength 1.5pt
\begin{picture}(170,70)(0,5)
\thicklines
\put (25,45){\framebox(20,10)[c]{Enc. 2}}
\put (25,65){\framebox(20,10)[c]{Enc. 1}}
\put (25,15){\framebox(20,10)[c]{Enc. $n$}}
\put (60,15){\framebox(20,60)[c]{$W$}}
\put (110,15){\framebox(30,10)[c]{Dec. $n$}}
\put (110,65){\framebox(30,10)[c]{Dec. 1}}
\put (110,45){\framebox(30,10)[c]{Dec. 2}}
\put (10,20){\vector(1,0){15}}
\put (10,50){\vector(1,0){15}}
\put (10,70){\vector(1,0){15}}
\put (45,20){\vector(1,0){15}}
\put (45,50){\vector(1,0){15}}
\put (45,70){\vector(1,0){15}}
\put (80,50){\line(1,0){15}}
\put (95,50){\circle*{2}}
\put (95,45){\line(0,1){25}}
\put (95,45){\line(0,-1){25}}
\put (95,70){\vector(1,0){15}}
\put (95,50){\vector(1,0){15}}
\put (95,20){\vector(1,0){15}}
\put (140,70){\vector(1,0){15}}
\put (140,50){\vector(1,0){15}}
\put (140,20){\vector(1,0){15}}
\put (10,22){\makebox(0,0)[bc]{$m_n$}}
\put (10,52){\makebox(0,0)[bc]{$m_2$}}
\put (10,72){\makebox(0,0)[bc]{$m_1$}}
\put (52.5,22){\makebox(0,0)[bc]{$\bu_n$}}
\put (52.5,52){\makebox(0,0)[bc]{$\bu_2$}}
\put (52.5,72){\makebox(0,0)[bc]{$\bu_1$}}
\put (87.5,52){\makebox(0,0)[bc]{$\bz$}}
\put (155,22){\makebox(0,0)[bc]{$\hat{m}_n$}}
\put (155,52){\makebox(0,0)[bc]{$\hat{m}_2$}}
\put (155,72){\makebox(0,0)[bc]{$\hat{m}_1$}}
\put (10,38){\makebox(0,0)[c]{$\vdots$}}
\put (35,38){\makebox(0,0)[c]{$\vdots$}}
\put (52.5,38){\makebox(0,0)[c]{$\vdots$}}
\put (102.5,38){\makebox(0,0)[c]{$\vdots$}}
\put (147.5,38){\makebox(0,0)[c]{$\vdots$}}
\put (115,30){\vector(0,-1){5}}
\put (135,30){\vector(0,-1){5}}
\put (125,30){\makebox(0,0)[c]{$\cdots$}}
\put (115,31){\makebox(0,0)[cb]{$\hbu_1$}}
\put (135,31){\makebox(0,0)[cb]{$\hbu_{n-1}$}}
\put (125,60){\vector(0,-1){5}}
\put (120,57){\makebox(0,0)[cb]{$\hbu_1$}}
\end{picture}
\caption{Channel splitting by multi-level coding.}
\label{gendecoder}
\end{center}
\end{figure}

The multi-level coding system here is designed around a random code ensemble for channel $W$,
specified by a random vector 
$\bU=(U_1,\ldots,U_n)\sim Q_1(x_1)\cdots Q_n(x_n)$ where $Q_i$ is a probability distribution on ${\cal X}$, $1\le i \le n$.
Intuitively, $U_i$ corresponds to the input random variable that is transmitted at the $i$th input terminal.
If we employ a sequential decoder that decodes the subchannels one at a time, applying successive cancellation between stages,
the sum cutoff rate can be as high as 
\begin{align*}
R_{0,S}(\bU, Z) \defn R_0(U_1,Z) +  \cdots + R_0(U_n,Z|U_1\cdots U_{n-1})
\end{align*}
where for any three random vectors $(U,V,Z)\sim P(u,v,z)$ 
\begin{align*}
R_{0}(U,Z|V) &\defn -\log \sum_{v} P(v) \sum_{z} \left[ \sum_{u} P(u|v) \sqrt{P(z|u,v)}\right]^2
\end{align*} 
This sum cutoff rate is to be compared with the ordinary cutoff rate $R_{0}(W) = \max_{Q} R_0(Q,W)$
where the maximum is over all $Q(u_1,\ldots,u_n)$, not necessarily in product-form.
A coding gain is achieved if $R_{0,S}(\bU,Z)$ is larger than $R_0(W)$.
Since $R_0(W)=nR_0(V)$ for all bijective label maps $f$, by the parallel-channels theorem mentioned earlier,
we may compare the normalized sum cutoff rate 
\begin{align*}
\hat{R}_{0,S}(\bU,Z) \defn \frac{1}{n}\,R_{0,S}(\bU,Z)
\end{align*}
with $R_0(V)$ to see if there is a coding gain.

The general framework described above admits a method by Pinsker \cite{Pinsker} 
that shows that if a sufficiently large number of copies of a DMC are combined, 
the sum cutoff rate can be made arbitrarily close 
to channel capacity.
Unfortunately, the complexity of Pinsker's scheme grows exponentially 
with the number of channels combined.
Although not practical, Pinsker's result is reassuring as far as the above method is concerned;
and, the main question becomes one of understanding how fast the sum cutoff rate 
improves as one increases the number of channels combined.

\section{BEC and BSC examples}\label{sec:examples}

\begin{figure}
\begin{center}
\resizebox{!}{!}{
\includegraphics*[width=\columnwidth]{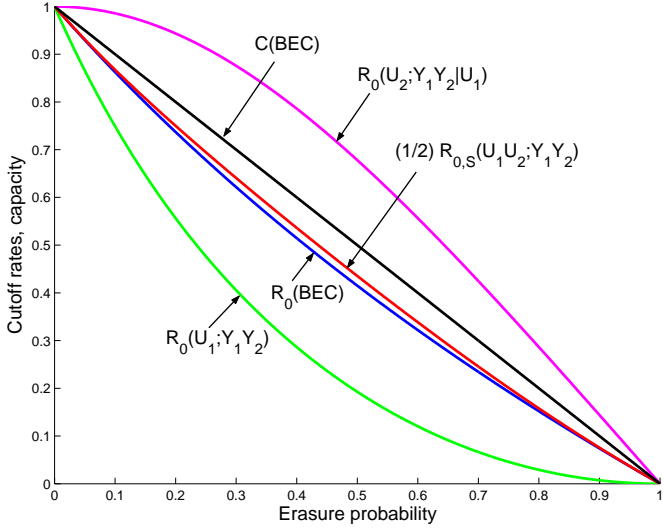}
}
\caption{Cutoff rates for the splitting of BEC.}
\label{r0compbec}
\end{center}
\end{figure}

The goal of this section is to illustrate the effectiveness of the abobe method by giving two examples,
where appreciable improvements in the cutoff rate are obtained by combining just two copies of a given channel.
\begin{Example}[BEC]\label{Ex:bec}
Let $V:{\cal X}\to {\cal Y}$ be a BEC with alphabets ${\cal X}=\{0,1\}$, ${\cal Y}=\{0,1,?\}$, and erasure probability $\epsilon$.
Consider combining two independent copies of $V$ 
to obtain a channel $W:{\cal X}^2\to {\cal Y}^2$ by means of the label map
\begin{align*}
f:(u_1,u_2)\to (x_1,x_2)=(u_1\oplus u_2, u_2)
\end{align*}
where $\oplus$ denotes modulo-2 addition.
Let the input variables be specified as $(U_1,U_2)\sim Q_1(u_1)Q_2(u_2)$ where $Q_1$, $Q_2$ are uniform on $\{0,1\}$.
Then, we compute that
\begin{align*}
R_0(U_1,Y_1Y_2) &= 1 - \log(1 + 2\epsilon -\epsilon^2)\\
R_0(U_2,Y_1Y_2|U_1) & =1 - \log(1 + \epsilon^2)
\end{align*}
An interpretation of these cutoff rates can be given by observing that user 1's channel, $u_1\to (y_1,y_2)$,
is effectively a BEC with erasure 
probability $1-(1-\epsilon)^2=2\epsilon - \epsilon^2$; an erasure occurs in this channel when either $x_1$ or $x_2$ is erased.
On the other hand, given that decoder 2 is supplied with the correct value of $u_1$, the channel seen by user 2 is a BEC with erasure probability 
$\epsilon^2$; an erasure occurs only when both $x_1$ and $x_2$ are erased.
The normalized sum cutoff rate under this scheme is given by
\begin{align*}
\hat{R}_{0,S}(U_1U_2,Y_1Y_2)= 1 - \frac{1}{2} \big[ \log(1 + 2\epsilon -\epsilon^2) + \log(1 + \epsilon^2) \big]
\end{align*}
which is to be be compared with the ordinary cutoff rate 
of the BEC, $R_0(V) =  1 - \log(1+ \epsilon)$.
These cutoff rates are shown in Fig.~\ref{r0compbec}.
The figure shows and it can be verified analytically that the above method improves the cutoff rate for all $0< \epsilon < 1$.
\end{Example}

\begin{figure}[hbt]
\begin{center}
\resizebox{!}{!}{
\includegraphics*[width=\columnwidth]{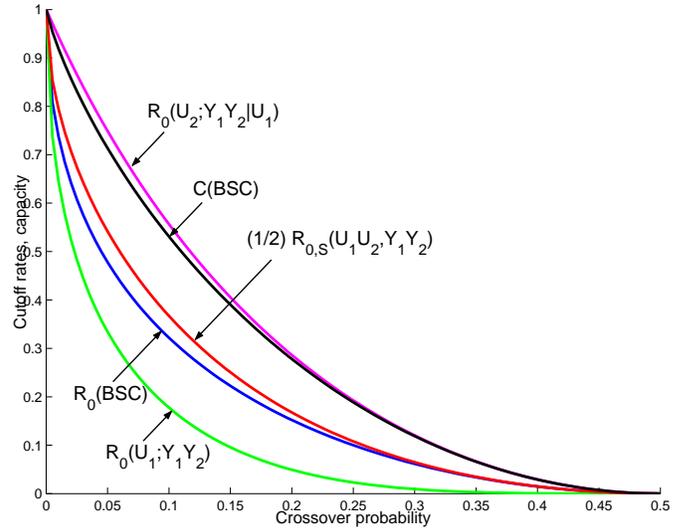}
}
\caption{Cutoff rates for the splitting of BSC.}
\label{r0compbsc}
\end{center}
\end{figure}

\begin{Example}[BSC]\label{Ex:bsc}
Let $V:{\cal X}\to {\cal Y}$ be a BSC with ${\cal X}={\cal Y}=\{0,1\}$ and crossover probability $0\le \epsilon \le 1/2$.
The cutoff rate of the BSC is given by
\begin{align*}
R_0(V) = 1 - \log (1 + \gamma(\epsilon))
\end{align*}
where $\gamma(\delta): = \sqrt{4 \delta (1-\delta)}$ for $0\le \delta\le 1$.

We combine two copies of the BSC using the label map $f:(u_1,u_2)\to (x_1,x_2)=(u_1\oplus u_2, u_2)$,
and take input variables $(U_1,U_2)\sim Q_1(x_1)Q_2(x_2)$ where $Q_1$, $Q_2$ are uniform on $\{0,1\}$.
The cutoff rates $R_0(U_1,Y_1Y_2)$ and $R_0(U_2,Y_1Y_2|U_1)$ can be obtained by direct calculation;
however, it is instructive to obtain them by the following argument.
The input and output variables of the channel $W$ are related by $y_1  = u_1 \oplus u_2 \oplus e_1$ and
$y_2  = u_2 \oplus e_2$
where $e_1$ and $e_2$ are independent noise terms, each taking the values 0 and 1 with probabilities $1-\epsilon$ and $\epsilon$, respectively.
Decoder 1 sees effectively the channel $u_1\to u_1\oplus e_1 \oplus e_2$,
which is a BSC with crossover probability $\epsilon_2 = 2 \epsilon(1-\epsilon)$ and
has cutoff rate 
\begin{align*} 
R_0(U_1,Y_1Y_2) =  1 - \log(1 + \gamma(\epsilon_2)) 
\end{align*}
Decoder 2 sees the channel $u_2 \to (y_1,y_2)$ and receives $u_1$ from decoder 1,
which is equivalent to the channel 
$u_2 \to (y_1\oplus u_1,y_2)=(u_2\oplus e_1,u_2\oplus e_2)$, 
which in turn is a BSC with diversity order 2 and has cutoff rate 
\begin{align*}
R_0(U_2,Y_1Y_2|U_1) = 1 - \log(1 + \gamma(\epsilon)^2)
\end{align*}
Thus, the normalized sum cutoff rate with this splitting scheme is given by
\begin{align*}
\hat{R}_{0,S}(U_1U_2,Y_1Y_2)= 1 - \frac{1}{2} \big[ \log(1 + \gamma(\epsilon_2)) + \log(1 + \gamma(\epsilon)^2) \big]
\end{align*}
which is larger than $R_0(V)$ for all $0<\epsilon < 0.5$, as shown in Fig.~\ref{r0compbsc}. 
\end{Example}

\section{Linear label maps}\label{sec:linearlabeling}
This section builds on the method employed in the previous section by considering general
types of {\sl linear} input maps.
Specifically, we consider combining $n$ independent copies of a BSC  
using a linear label map $\bx=\bu F$ where $F$ is an invertible matrix of size $n\times n$.
The channel output is given by $\by = \bx +\be$ where $\be$ is the noise vector.
Throughout, we use an input ensemble $\bU=(U_1,\ldots,U_n)$ consisting of i.i.d. components, each
component equally likely to take the values 0 and 1.
In the rest of this section, we give two methods that follow this general idea.
\subsection{Kronecker powers of a given labeling}
We consider here linear maps of the form $F=A^{\otimes k}$ where 
$A=\bigl[ \begin{smallmatrix} 1 & 0 \\ 1 & 1 \end{smallmatrix} \bigr]$
is the linear map used in Ex.~\ref{Ex:bsc}.
The normalized sum cutoff rates for such $F$ are 
listed in the following table for a BSC with error probability of $\epsilon=0.1$.
The cutoff rate and capacity of the same BSC are $R_0=.3219$ and $C=.5310$.
\begin{center}
\begin{tabular}{c|cccc}\hline
$k$ & 1 & 2 & 3 & 4\\\hline
$\hat{R}_{0,S}$ & .3670 & .4016 & .4245 & .4433\\ \hline
\end{tabular} 
\end{center} 
The scheme with $F_k$ has $n=2^k$ subchannels and 
the size of the output alphabet of the combined channel equals $2^{2^k}$.
The rapid growth of this number prevented computing $\hat{R}_{0,S}$ for $k\ge 5$.
\subsection{Label maps from block codes}
Let $G=[\,P \;\;I_{k}\,]$ be the generator matrix in systematic form of a $(n,k)$ linear binary block code ${\cal C}$.
Here, $P$ is a $k \times (n-k)$ matrix and $I_k$ is the $k$-dimensional identity matrix.
A linear label map is obtained by setting  
\begin{align}\label{blockcodeF}
F  = \left[ 
\begin{array}{c|c}
I_{n-k} & 0 \\
\hline
P & I_{k}
\end{array}
\right]
\end{align}
Note that $F^{-1}=F$ and
that the first $(n-k)$ columns of $F$ equals $H^T$, the tranpose of a {\sl parity-check} matrix for
${\cal C}$.
Thus, when the receiver computes the vector $\bv = \by F^{-1} = \by F$, the first $(n-k)$ coordinates of $\bv$ have the form
$v_i = u_i \oplus s_i$, $1\le i \le n-k$,
where $s_i$ is the $i$th element of the {\sl syndrome} vector $\bs = \by H^T= \be H^T$.
This $i$th ``syndrome subchannel'' is effectively the cascade of $k$ BSCs (each with crossover probability $\epsilon$) where $k$ is the number
of 1's in the $i$th row of $H$.
The remaining subchannels, which we call ``information subchannels,'' have the form
$v_i = u_i \oplus e_i$, $(n-k+1)\le i \le n$.
\begin{Example}[Dual of Golay code]\label{Ex:dualGolay} Let $F$ be as in \eqref{blockcodeF} with $n=23$, $k=11$, and 
\begin{align*}
P= \left[
	\begin{array}{cccccccccccc}
	1&0&0&1&1&1&0&0&0&1&1&1\\
      1&0&1&0&1&1&0&1&1&0&0&1\\
      1&0&1&1&0&1&1&0&1&0&1&0\\
      1&0&1&1&1&0&1&1&0&1&0&0\\
      1&1&0&0&1&1&1&0&1&1&0&0\\
      1&1&0&1&0&1&1&1&0&0&0&1\\
      1&1&0&1&1&0&0&1&1&0&1&0\\
      1&1&1&0&0&1&0&1&0&1&1&0\\
      1&1&1&0&1&0&1&0&0&0&1&1\\
      1&1&1&1&0&0&0&0&1&1&0&1\\
      0&1&1&1&1&1&1&1&1&1&1&1
      \end{array}
	\right]
\end{align*}
The code with the generator matrix $G=[\,P\;I_{11}\,]$ is the dual of the Golay code \cite[p.~119]{Blahut83}.
We computed the normalized sum cutoff rate $\hat{R}_{0,S}=.4503$ at $\epsilon=0.1$ for this scheme.
The rate allocation vector $(R_0(U_i;\bY|U_1,\ldots,U_{i-1}):1\le i\le 23)$ is shown in Fig.~\ref{Fig:Golay}.
There is a jump in the rate allocation vector in going from the syndrome subchannels to information subchannels, as expected.
\begin{figure}[hbt]
\begin{center}
\resizebox{!}{!}{
\includegraphics*[width=\columnwidth]{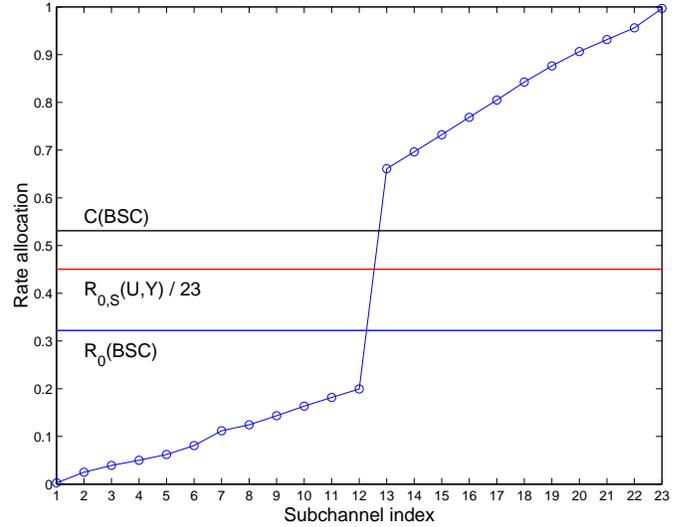}
}
\caption{Rate allocation for Ex.~\ref{Ex:dualGolay}.}
\label{Fig:Golay}
\end{center}
\end{figure}
\end{Example}

\section{Concluding remarks}\label{sec:conclusions}
We have presented a method for improving the sum cutoff rate of a given DMC based on channel combining and splitting.
Although the method has been presented for some binary-input channels, 
it is readily applicable to a wider class of channels.
Our starting point for studying this problem is rooted in the literature on methods to improve the cutoff rate in
sequential decoding, most notably, Pinsker's \cite{Pinsker} and Massey's \cite{Massey81} works;
however, the method we proposed has many common elements with well-known coded-modulation techniques, namely,
Imai and Hirakawa's \cite{Imai77} multi-level coding scheme
and Ungerboeck's \cite{Ungerboeck1} set-partioning idea, 
which corresponds to the relabeling of inputs in our approach.
In this connection, we should cite the paper by Wachsmann et al \cite{Wachsmann99} which develops
design methods for coded modulation using the sum cutoff rate and random-coding exponent as figures of merit.

Our main aim has been to explore the existence of practical
schemes that boost the sum cutoff rate to near channel capacity.
This goal remains only partially achieved.
Further work is needed to understand if this is a realistic goal.
\bibliographystyle{ieeetr}
\bibliography{references}

\end{document}